\documentclass{iopart}

\usepackage[dvips]{graphicx}

\begin{document}

\title[Photon round trips in the field of a plane gravitational wave]
{On the round-trip time for a photon propagating in the field of a
plane gravitational wave}

\author{M~Rakhmanov}
\address{Department of Physics and Astronomy, 
University of Texas at Brownsville, Brownsville, TX 78520, USA}

\ead{malik.rakhmanov@utb.edu}

\begin{abstract}
A network of large-scale laser interferometers is currently employed 
for searches of gravitational waves from various astrophysical 
sources. The frequency dependence of the dynamic response of these 
detectors introduces corrections to their antenna patterns which in 
principle can affect the outcome of the associated data-analysis 
algorithms. The magnitude of these corrections and the corresponding 
systematic errors have recently been estimated for searches 
of periodic and stochastic gravitational waves 
(2008 {\it Class.~Quantum~Grav.} {\bf{25}}~184017). However, the 
calculation of the detector response in that paper followed the 
traditional semi-rigorous approach which does not properly take into 
account the curved nature of spacetime. The question then arises as 
to whether the results will be the same if the calculation is done 
within the rigorous framework of general relativity. In this paper 
we provide such a derivation of the response of the detectors to 
gravitational waves. We obtain the photon propagation time from the 
solution of the equation for null geodesics and calculate the 
corresponding phase delay by solving the eikonal equation for 
curved spacetime. The calculations are then extended to include 
phase amplification from multi-beam interference in Fabry-Perot 
resonators which play an important role in the formation of the 
signal in these detectors.
\end{abstract}


\section{Introduction}

Kilometer-scale gravitational wave detectors such as LIGO
\cite{Barish:1999} and VIRGO \cite{Bradaschia:1990} have recently
reached their design sensitivities and have already established a
number of astrophysical upper limits. The associated data-analysis
efforts are currently focused on the development of algorithms for
accurate waveform reconstruction and parameter estimation of the
anticipated signal. The accuracy of these algorithms depends on the
models which are utilized to describe the detector response to
gravitational waves. In particular, the long-wavelength approximation
\cite{Thorne:1987} commonly practiced in such models introduces
systematic errors in parameter estimation \cite{Baskaran:2004} which
at first were thought to be large enough to modify the outcome of 
search algorithms, and even the reported upper limits. Recent studies, 
however, have shown that these errors were previously over-estimated; 
and, in fact, they do not significantly affect the search algorithms 
at the frequencies considered thus far 
\cite{Rakhmanov:2008}. Curiously, in the course of these studies, it 
was realized that the usual derivations of the detector response 
(see, e.g. \cite{Gursel:1984, Christensen:1992, Sigg:1997, 
  Schilling:1997, Larson:2000, Cornish:2001, Cornish:2003} and more 
recently \cite{Baskaran:2004, Rakhmanov:2008}) only partially follow 
the principles of general relativity. For example, in deriving the 
photon round-trip time, one usually integrates metric perturbations 
along the unperturbed photon trajectory, assuming that it is a 
straight line, and thus neglecting the slight bending of the 
trajectory due to the gravitational wave \cite{Whelan:2007}. The 
calculations of the phase acquired by a photon in one round trip are 
also carried out assuming an idealized electromagnetic wave 
propagation, i.e. along the straight line. Such assumptions were also 
made in the recent analysis of the LIGO antenna patterns 
\cite{Rakhmanov:2006, Rakhmanov:2008}. In this paper we resolve these 
problems by providing a fully-relativistic analysis which properly 
takes into account the curved nature of the spacetime. We derive the 
photon propagation time by solving the equation for null 
geodesics. The round-trip phase is then found by solving the equation 
for eikonal in curved spacetime. The results agree with those obtained 
within the semi-rigorous approach.

\section{Equation for null geodesics}
\label{propLight}

Consider spacetime in which a plane-fronted gravitational wave is 
propagating on a flat background. Let the coordinates of this
spacetime be $x^{\mu}=(ct, x, y, z)$ with Greek suffixes taking values 
$0,1,2,3$. Within the linearized\footnote{The analysis in this paper 
belongs to the linearized theory and therefore all the calculations
are valid to first order in $h$ only.} theory, the metric of this
spacetime is given by
\begin{equation}\label{linMetric}
   g_{\mu \nu} = \eta_{\mu \nu} + h_{\mu \nu} ,
\end{equation}
where $\eta_{\mu \nu} = {\mathrm{diag}}\{-1,1,1,1\}$ is the 
Minkowskii tensor and $h_{\mu\nu}$ is a small perturbation which
represents the gravitational wave. In the transverse traceless gauge 
\cite{Misner:1973} the metric perturbation takes a particularly
simple form:
\begin{equation}
   h_{\mu \nu} = \left[
   \begin{array}{ccrc}
    0    &              &               &    \\
         &  h_{+}       &   h_{\times}  &    \\
         &  h_{\times}  &  -h_{+}       &    \\
         &              &               &  0
   \end{array}\right] ,
\end{equation}
where $h_{+}$ and $h_{\times}$ are functions of $ct + z$ only. Here we 
assumed that the gravitational wave is propagating in the negative $z$ 
direction and its principal axes of polarization are chosen along the 
$x$ and $y$ coordinates. We will also use two auxiliary coordinates,
\begin{equation}
   u = ct + z
   \qquad {\mathrm{and}} \qquad 
   v = ct - z ,
\end{equation}
to simplify some of the following calculations. The fundamental form 
corresponding to metric (\ref{linMetric}) is given by
\begin{equation}\label{metricForm2}
   ds^2 = - du \, dv + dx^2 + dy^2 + h_{+}(u) (dx^2 - dy^2) + 
           2 h_{\times}(u) \, dx \, dy .
\end{equation}

Propagation of light in this spacetime is described by a null
geodesic: $x^{\mu}=x^{\mu}(\sigma)$, where $\sigma$ is the affine 
parameter along the curve. The tangent vector to this curve, which is
also the photon momentum, 
\begin{equation}
   p^{\mu} = \frac{d x^{\mu}}{d \sigma} 
   \qquad {\mathrm{and}} \qquad 
   p_{\mu} =  g_{\mu\nu} \, p^{\nu} ,
\end{equation}
satisfies the null condition:
\begin{equation}\label{nullCond}
   p_{\mu} p^{\mu} = 0 .
\end{equation}
In the explicit form, the covariant components of the tangent vector
are 
\begin{eqnarray}
   p_v & = & -\frac{1}{2} \frac{du}{d\sigma}  , \label{def_pv}\\
   p_x & = & \left[ 1 + h_{+}(u) \right] \frac{dx}{d\sigma} + 
             h_{\times}(u) \frac{dy}{d\sigma} , \label{def_px}\\
   p_y & = & \left[ 1 - h_{+}(u) \right] \frac{dy}{d\sigma} + 
             h_{\times}(u) \frac{dx}{d\sigma} , \label{def_py}\\
   p_u & = & -\frac{1}{2} \frac{dv}{d\sigma}  . \label{def_pu} 
\end{eqnarray}
The geodesic equation can be written as
\begin{equation}
   \frac{d p_{\alpha}}{d \sigma} = 
      \frac{1}{2} \, g_{\mu\nu, \alpha} \, p^{\mu} p^{\nu} .
\end{equation}
Since the metric does not depend on the $x$, $y$, and $v$ coordinates,
the corresponding components of the tangent vector are constant:
\begin{eqnarray}
   p_v & = & p_{v0} , \label{eqForPv} \\
   p_x & = & p_{x0} , \label{eqForPx} \\
   p_y & = & p_{y0} . \label{eqForPy} 
\end{eqnarray}
The fourth equation is
\begin{equation}\label{eqForPu}
   \frac{d p_u}{d \sigma} = \frac{1}{2} 
      \left( p_x^2 - p_y^2 \right) h'_{+}(u) + p_x p_y \, h'_{\times}(u) . 
\end{equation}
Its solution can be obtained directly from the null condition 
(\ref{nullCond}), and yields
\begin{equation}
   p_{u}(u) = \frac{1}{4 p_{v0}} \left[ p_{x0}^2 +
           p_{y0}^2 - (p_{x0}^2 - p_{y0}^2)h_{+}(u) - 
           2 p_{x0} p_{y0} h_{\times}(u) \right] .
           \label{puFromNorm} 
\end{equation}
This completes the first integration of the geodesic equation.

Next, equation (\ref{eqForPv}) together with the definition for $p_v$
in (\ref{def_pv}) can be integrated and yields
\begin{equation}\label{eqForU(s)}
   u(\sigma) = u_0 - 2 p_{v0} \sigma ,
\end{equation}
where $u_0$ is the initial value for this coordinate. Equations 
(\ref{eqForPx}) and (\ref{eqForPy}) together with definitions 
(\ref{def_px}) and (\ref{def_py}) yield the solution for $x$ and $y$:
\begin{eqnarray}
   x(\sigma) & = & x_0 + p_{x0} \sigma [1 - f_{+}(\sigma)] - 
              p_{y0} \sigma \, f_{\times}(\sigma),
              \label{eqForX(s)} \\
   y(\sigma) & = & y_0 + p_{y0} \sigma [1 +  f_{+}(\sigma)] - 
              p_{x0} \sigma \, f_{\times}(\sigma) ,
              \label{eqForY(s)} 
\end{eqnarray}
where $x_0$ and $y_0$ are the initial values for these coordinates, and
$f_i$ are the {\emph{averaged}} amplitudes of the gravitational wave:
\begin{equation}\label{gFactor}
   f_i(\sigma) = \frac{1}{\sigma} \int_0^{\sigma} 
      h_{i}[u(\sigma')] d\sigma' 
\end{equation}
for $i = +, \times$. Finally, from (\ref{def_pu}) and
(\ref{puFromNorm}) we obtain the solution for $v$:
\begin{equation}\label{v(sigma)}
   v(\sigma) = v_0 - 2 \int_0^{\sigma} p_u[u(\sigma')] \, d\sigma' ,
\end{equation}
where $v_0$ is the initial value for this coordinate. In the explicit
form it is given by
\begin{eqnarray}\label{eqForV(s)}
   v(\sigma) & = & v_0 - \frac{\sigma}{2 p_{v0}} 
                   \left( p_{x0}^2 + p_{y0}^2 \right) - \nonumber \\
             &   & \frac{\sigma}{2 p_{v0}} \left[ 
                   (p_{x0}^2 - p_{y0}^2) f_{+}(\sigma) +
                   2 \, p_{x0} p_{y0} \; f_{\times}(\sigma) \right] .
\end{eqnarray}
This completes the second integration of the geodesic equation. 
The arbitrary constants: $p_{v0}$, $p_{x0}$, $p_{y0}$, 
$u_0$, $x_0$, $y_0$, and $v_0$ will be determined from the boundary 
conditions.

\section{Round-trip propagation time}
\subsection{Boundary-value problem}
\label{bvalueProb}

Consider a photon trajectory with fixed boundaries: $P_0=(0,0,0)$ and 
$P=(x,y,z)$, and assume that it starts at $P_0$ at time $t_0$ and ends 
at $P$ at time $t$. The trajectory is given by the null geodesic 
described in (\ref{eqForU(s)}--\ref{eqForY(s)})
and (\ref{eqForV(s)}) with the following initial values:
$x_0 = y_0 = z_0 = 0$ and $u_0 = v_0 = c t_0$. 
The first three constants of integration can be found from 
(\ref{eqForU(s)}), (\ref{eqForX(s)}), and (\ref{eqForY(s)}):
\begin{eqnarray}
   p_{v0} & = & - \frac{u - u_0}{2 \sigma} , \label{pv0} \\
   p_{x0} & = & (1 + f_{+}) \frac{x}{\sigma} + 
              f_{\times} \, \frac{y}{\sigma}, \label{px0} \\
   p_{y0} & = & (1 - f_{+}) \frac{y}{\sigma} + 
              f_{\times} \, \frac{x}{\sigma}. \label{py0}
\end{eqnarray}
In these equations $\sigma$ stands for the value of the affine
parameter at the end point, $P$. Substituting (\ref{px0}) and
(\ref{py0}) into (\ref{eqForV(s)}), we obtain:
\begin{equation}
   v = c t_0 - \frac{1}{2 p_{v0} \sigma} \left[
   x^2 (1 + f_{+}) + y^2 (1 - f_{+}) + 2 x y f_{\times} \right] ,
\end{equation}
in which we can replace $p_{v0}$ from (\ref{pv0}) with the result
\begin{equation}\label{eqForDT1}
   c^2 (t - t_0)^2 = r^2 + (x^2 - y^2) f_{+} + 2 x y f_{\times} .
\end{equation}
Here we introduced the short-hand notation: 
$r \equiv \sqrt{x^2 + y^2 + z^2}$.

If upon rearching point $P$, the photon changes its course and
proceeds along a different trajectory the solution can be
found in a similar way. Consider for example the situation when the
photon travels back to the point of origin. In other words, assume 
that the photon trajectory starts at $P$ at time $t_1$ and ends at 
$P_0$ at time $t$. In this case, the trajectory is also given by 
equations (\ref{eqForU(s)}--\ref{eqForY(s)}) and (\ref{eqForV(s)}) 
but with different boundary conditions. The initial values for the 
coordinates are now $ct_1, x, y, z$, whereas the final values are 
$x = y = z = 0$ and $u = v = c t$. Consequently, the first three 
constants of integration are:
\begin{eqnarray}
   q_{v0} & = & - \frac{u - u_0}{2 \sigma} , \label{qv0} \\
   q_{x0} & = & - (1 + g_{+}) \frac{x}{\sigma} - 
              g_{\times} \, \frac{y}{\sigma}, \label{qx0} \\
   q_{y0} & = & - (1 - g_{+}) \frac{y}{\sigma} - 
              g_{\times} \, \frac{x}{\sigma}, \label{qy0}
\end{eqnarray}
where $g_i$ are the averaged amplitudes of the gravitational wave, 
(\ref{gFactor}), for the return trip. Substituting equations (\ref{qv0}),
(\ref{qx0}), and (\ref{qy0}) into (\ref{eqForV(s)}), we obtain 
\begin{equation}\label{eqForDT2}
   c^2 (t - t_1)^2 = r^2 + (x^2 - y^2) g_{+} + 2 x y g_{\times}.
\end{equation}
Equations (\ref{eqForDT1}) and (\ref{eqForDT2}) determine the duration 
for the forward and the return trips.

\subsection{Photon propagation time}

We can now apply the above formulae to calculate the round-trip time 
for a photon propagating between two inertial test masses. Assume that 
the points $P_0$ and $P$ correspond to the two test masses. Note that 
the coordinates of the test masses do not change under the influence 
of a gravitational wave \cite{Misner:1973}, so that this physical 
situation indeed corresponds to the solution of the geodesic equation 
with fixed boundaries described above. Consider the equation for the 
forward propagation time (\ref{eqForDT1}). To first order in $h$, its 
solution is 
\begin{equation}\label{propTime1h}
   c(t - t_0) = r + \frac{1}{2 r} 
      \left[ (x^2 - y^2) f_{+} + 2 x y f_{\times}\right] .
\end{equation}
Let the unperturbed propagation time between the test masses be $T$,
i.e. $T = r/c$. Then the duration of the forward trip in the presence
of a gravitational wave is 
\begin{equation}
   T_1(t) = T + \frac{1}{2} T  
      \left[ (a_x^2 - a_y^2) f_{+}(t) + 2 a_x a_y f_{\times}(t) \right],
\end{equation}
where ${\hat{a}} \equiv (x/r,y/r,z/r)$.
Here we explicitly indicated that $f_i$ are functions of time. To find 
this functional dependence, we change the integration variable 
$\sigma'$ in (\ref{gFactor}) to $u'$, using (\ref{pv0}):
\begin{equation}
   f_i(t) = \frac{1}{u - u_0} \int_{u_0}^{u} h_{i}(u') \, \rmd u' .
\end{equation}
Next, we substitute $u' = u_0 + \alpha$, and then replace $u$ with its 
first order approximation, which can be found from 
$u - u_0 \approx r + z$. The result is 
\begin{equation}\label{gFactorA}
   f_i(t) = \frac{1}{r + z} \int_{0}^{r + z} 
      h_{i}(ct - r + \alpha) \, \rmd \alpha .
\end{equation}
Similar calculations can be carried out for the return trip and lead
to 
\begin{equation}
   T_2(t) = T + \frac{1}{2} T  
      \left[ (a_x^2 - a_y^2) g_{+}(t) + 2 a_x a_y g_{\times}(t) \right],
\end{equation}
except that in this case $u - u_0 \approx r - z$. Therefore,
\begin{equation}\label{gFactorB}
   g_{i}(t) = \frac{1}{r - z} \int_{0}^{r - z} 
      h_{i}(ct - r + z + \beta) \, \rmd \beta.
\end{equation}
The duration of the photon round trip can then be found as  
\begin{eqnarray}
   T_{\mathrm{r.t.}}(t) & = & T_1[t - T_2(t)] + T_2(t) \\
      & \approx & T_1(t - T) + T_2(t) .
\end{eqnarray}
It is worthwhile to separate the large unperturbed value from this 
quantity, and write the result as
\begin{equation}\label{TrtSum}
   T_{\mathrm{r.t.}}(t) = 2 T + \delta T_{\mathrm{r.t.}}(t) .
\end{equation}
The second term in the right-hand side represents the small variation
of the round-trip time which is induced by the gravitational wave,
\begin{equation}\label{TrtViaC}
   \delta T_{\mathrm{r.t.}}(t) = \frac{1}{2} \, T \left[  
      (a_x^2 - a_y^2) A_{+}(t) + 2 a_x a_y A_{\times}(t) \right] ,
\end{equation}
where 
\begin{equation}\label{gFactorC}
   A_{i}(t) = f_i(t - T) + g_i(t) .
\end{equation}
Further calculations become somewhat simpler if we switch to the
frequency domain by means of either Fourier or Laplace
transformations.

\subsection{Laplace-domain transfer function}

The Laplace transform of an arbitrary function of time $F(t)$ is
given by
\begin{equation}
   \tilde{F}(s) = \int_0^{\infty} \rme^{-st} F(t) \rmd t .
\end{equation}
Since $A_i$ are linear (integral) transforms of $h_i$, we can
introduce a transfer function $D(s,\hat{a})$ in the Laplace domain
such that 
\begin{equation}
   \tilde{A}_i(s) = D(s,\hat{a}) \tilde{h}_i(s) 
\end{equation}
for both $i = +$ and $\times$. In this equation we used the following 
definition: 
\begin{equation}
   \tilde{h}_{i}(s) = \int_0^{\infty} \rme^{-st} h_i(ct) \rmd t .
\end{equation}
Then (\ref{TrtViaC}) can be written as 
\begin{equation}\label{TrtDs}
   \delta \tilde{T}_{\mathrm{r.t.}}(s) = 
      T D(s,\hat{a}) \left[ (a_x^2 - a_y^2) \tilde{h}_{+}(s) + 
      2 a_x a_y \tilde{h}_{\times}(s) \right] ,
\end{equation}
where $D(s,\hat{a})$ is the corresponding transfer function:
\begin{equation}\label{tfD(s,a)}
   D(s,\hat{a}) = \frac{1}{2s T} \left[
      \frac{1 - \rme^{-(1 - a_z) sT}}{1 - a_z} - \rme^{-2sT} 
      \frac{1 - \rme^{ (1 + a_z) sT}}{1 + a_z} \right] .
\end{equation}
The Fourier-domain version of (\ref{tfD(s,a)}) can be obtained by the 
substitution, $s = 2\pi \rmi f$, and yields equation (14) in 
\cite{Rakhmanov:2008}. (For complete equivalence, one must also replace 
$a_z$ with $\hat{a} \cdot \hat{n}$, where $\hat{n}$ is the unit vector 
in the direction of the source of the gravitational waves.)

\section{Eikonal equation}
\label{eikonalEq}

A continuous electromagnetic wave with frequency $\omega$ and 
wavenumber $k$ ($k\equiv\omega/c$) is described by a scalar function 
$\phi(x)$ called the eikonal or phase. In this picture, the wavefronts 
are defined by the surfaces of constant phase and the rays, along which 
they propagate, are defined by the normals to these surfaces, 
\begin{equation}\label{defRays}
   p_{\mu} = - \frac{\partial \phi}{\partial x^{\mu}} .
\end{equation}
Consequently, the phase must satisfy the eikonal equation
\cite{Landau:1971}: 
\begin{equation}\label{eikonal}
   g^{\mu\nu} \frac{\partial \phi}{\partial x^{\mu}} 
              \frac{\partial \phi}{\partial x^{\nu}} = 0 .
\end{equation}
The solution can be found from (\ref{defRays}) by direct integration:
\begin{equation}\label{defPsi}
   \phi = \phi_0 - \int_{P_0}^{P} p_{\mu} \rmd x^{\mu} ,
\end{equation}
and in general is a function of two points, $P_0$ and $P$. Here
$\phi_0$ is the value of the eikonal at $P_0$ which may not
necessarily be constant.

If the ray originates at the source of the 
electro-magnetic waves, the first term in the right-hand side 
of (\ref{defPsi}) represents the phase of the source, 
$\phi_0 \equiv \phi_S(t_0)$, where 
\begin{equation}
   \phi_S(t) = \omega t .
\end{equation}
The second term in (\ref{defPsi}) represents the change of the phase 
along the ray,
\begin{equation}\label{DeltaPhi}
   \Delta \phi \equiv -
      \int_{P_0}^{P} p_{\mu} \rmd x^{\mu} .
\end{equation}
Since three out of four components of the photon momentum are
constant along the ray, we readily obtain
\begin{equation}
   \Delta \phi = - p_{x0} x - p_{y0} y - p_{v0} (v - v_0) -
      \int_{u_0}^{u} p_{u}(u') \rmd u' .
\end{equation}
For simplicity, we assumed that the coordinates of $P_0$ are 
$x_0 = y_0 = z_0 = 0$. 
Furthermore, the last two terms on the right-hand side of
(\ref{DeltaPhi}) are equal. This can be seen from (\ref{v(sigma)}) 
by changing the integration variable $\sigma$ to $u$: 
\begin{equation}
   v = v_0 + \frac{1}{p_{v0}} \int_{u_0}^{u} p_{u}(u') \rmd u' .
\end{equation}
Knowing this, we can write the phase change along the ray as 
\begin{equation}\label{dphiViaP}
   \Delta \phi = - p_{x0} x - p_{y0} y - 2 p_{v0} (v - v_0) .
\end{equation}
In the boundary-value problem considered in section \ref{bvalueProb}, 
$p_{\mu}$ are functions of the coordinates of the end point, 
(\ref{pv0}--\ref{py0}). Substituting these formulae into
(\ref{dphiViaP}), we obtain
\begin{equation}\label{dphiViaX}
   \Delta \phi(t_0,t) = \frac{1}{\sigma} \left[ c^2 (t - t_0)^2 - 
      r^2 - (x^2 + y^2) f_{+}(t) - 2 x y f_{\times}(t) \right] ,
\end{equation}
where we show explicitly its dual time dependence. 
Note that if we follow a surface of constant phase, the two points 
$P$ and $P_0$ will correspond to the same wavefront and therefore 
$\Delta \phi_1 = 0$. In this case, (\ref{dphiViaX}) will yield the 
photon propagation time (\ref{eqForDT1}).

\section{Round-trip phase}
\label{rtPhase}

Consider now a situation when the wave is reflected at some point, say
$P_1$, and from then on proceeds along a different ray. Let the end
point along the new ray be $P$. Since the eikonal is an additive quantity, 
its value at the end point is
\begin{equation}\label{phaseAccum}
   \phi(t) = \phi_0(t_0) + \Delta \phi_1(t_0,t_1) + \Delta \phi_2(t_1,t) .
\end{equation}
Here $\Delta \phi_1$ is given by (\ref{dphiViaX}) with $t_1$ replacing
$t$ and 
\begin{equation}\label{netPhase2}
   \Delta \phi_2 = - \int_{P_1}^{P} q_{\mu} \rmd x^{\mu} ,
\end{equation}
where $q_{\mu}$ is the photon momentum along the second ray. 
Obviously, this approach can be extended to an arbitrary number of
reflections.

We can now apply (\ref{phaseAccum}) to find the phase of the wave
returning to the source after one round trip. In this case $q_{\mu}$ 
are given by (\ref{qv0}--\ref{qy0}). Sustituting these formulae into 
(\ref{netPhase2}), we obtain the phase increment for the return trip,
\begin{equation}\label{deltaPhi2}
   \Delta \phi_2(t_1, t) = \frac{1}{\sigma} \left[ c^2 (t - t_1)^2 - 
      r^2 - (x^2 + y^2) g_{+}(t) - 2 x y g_{\times}(t) \right] .
\end{equation}
In this case too, the condition for constant phase along the ray 
($\Delta \phi_2 = 0$) yields the photon propagation time
(\ref{eqForDT2}).

Following a particular wave along the round trip, we find that both
phase increments $\Delta \phi_1$ and $\Delta \phi_2$ vanish and that 
the phase at the end of the round trip is 
\begin{equation}
   \phi(t) = \phi_S(t_0) = \omega t_0 ,
\end{equation}
in which $t_0$ lags $t$ by the duration of one round trip:
\begin{equation}\label{phaseAfterRT}
   \phi(t) = \omega [t - T_{\mathrm{r.t.}}(t)] .
\end{equation}
Thus, the phase difference between the wave returning to the source 
and the wave presently emmited by the source is 
\begin{equation}
   \phi(t) - \phi_S(t) = -\omega T_{\mathrm{r.t.}}(t) = 
      - 2kL - \psi(t) ,
\end{equation}
where $k L = \omega T$ is the large unperturbed value of the phase, and 
\begin{equation}\label{psiDef}
   \psi(t) = \omega \delta T_{\mathrm{r.t.}}(t) 
\end{equation}
is the small phase shift due to the presence of the gravitational
wave. We have thus proved equation (15) in \cite{Rakhmanov:2008}.

\section{Light storage in a Fabry-Perot resonator}
\subsection{Superposition of fields}

\begin{figure}
 \centering\includegraphics[width=\columnwidth]{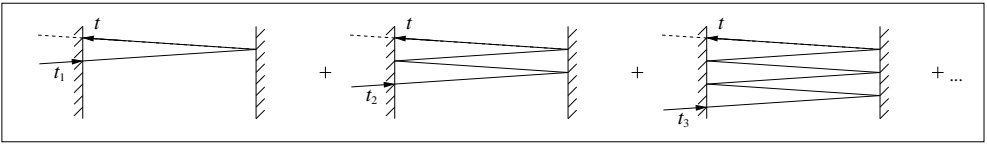}
   \caption{Schematic representation of the superposition of fields 
   in a Fabry-Perot resonator. At any given time, $t$, the complex 
   amplitude of the wave inside the resonator can be viewed as a sum 
   of the complex amplitudes of an infinite number of waves. Each wave 
   corresponds to a ray which completed a certain number of round 
   trips: $1, 2, 3, \ldots, n$. The rays enter the resonator at times 
   $t_1, t_2, t_3, \dots t_n$.}
   \label{multipleBeams}
\end{figure}

The phase shift acquired by the light during one round trip can be 
amplified by means of multiple reflections. One way of doing so is to 
store the light in a Fabry-Perot resonator which consists of two 
mirrors separated by a distance $L$ from each other. The analysis of 
single wave propagation in sections \ref{eikonalEq} and \ref{rtPhase} 
can be extended to multiple waves using the principle of superposition. 
Namely, the wave inside a Fabry-Perot resonator can be viewed as a 
superposition of waves corresponding to rays which have undergone multiple 
reflections inside the resonator as shown in figure \ref{multipleBeams}.

Consider a wave incident on a Fabry-Perot resonator and assume that 
its amplitude is constant in time, $A_0$. As we know, the phase of 
the wave will evolve according to the eikonal equation 
(\ref{eikonal}) and in general will be a function of position and 
time. In what follows, all quantities will be considered at a fixed 
location (the front mirror of the resonator) and therefore will be 
functions of time only. Then the complex amplitude of the wave 
incident on the front mirror of the resonator can be written as 
${\mathcal{E}}_{\mathrm{in}}(t) = A_0 \rme^{\rmi \phi(t)}$. 
The complex amplitude of the electromagnetic wave inside the 
Fabry-Perot resonator can be found by superposition:
\begin{equation}\label{superPos}
   {\mathcal{E}}(t) \equiv \sum_{n=1}^{\infty} {\mathcal{E}}_n(t) = 
     \sum_{n=1}^{\infty} A_n \rme^{\rmi \phi_n(t)} ,
\end{equation}
where $A_n$ and $\phi_n(t)$ are the amplitude and phase of the
wave, which corresponds to the $n$-th ray inside the resonator. 
Let the reflectivity of the mirrors be $r$ and $r'$ and the 
transmissivity of the front mirror be $\tau$. Then 
$A_1 = A_0 \tau (-r')$ and 
\begin{equation}
   A_n = A_1 (r r')^{n-1} = - \frac{\tau}{r} A_0 (r r')^{n} . 
\end{equation}
(Here we assumed the usual sign reversal upon reflection.) 
The phases which correspond to the individual rays can be found from 
(\ref{phaseAfterRT}). Namely, the phase of the first ray is 
\begin{equation}
   \phi_1(t) = \phi_S(t_1), 
   \quad {\mathrm{where}} \quad 
   t_1 = t - T_{\mathrm{r.t.}}(t) ,
\end{equation}
the phase of the second ray is
\begin{equation}
   \phi_2(t) = \phi_S(t_2),
   \quad {\mathrm{where}} \quad 
   t_2 = t_1 - T_{\mathrm{r.t.}}(t_1) ,
\end{equation}
and so on. The general formula for the phase is 
\begin{equation}
   \phi_n(t) = \phi_S(t_n),
   \quad {\mathrm{where}} \quad 
   t_n = t - \sum_{m=0}^{n-1} T_{\mathrm{r.t.}}(t_m) ,
\end{equation}
for $n = 1, 2, 3, \ldots$.\footnote{Whenever $t_0$ occurs in this 
series it should be replaced with $t$.}
Separating $T_{\mathrm{r.t.}}(t)$ into two components, as in
(\ref{TrtSum}), and neglecting the terms of order $h$ in the argument of
$\delta T_{\mathrm{r.t.}}(t_m)$, we obtain $t_n$ in closed form:
\begin{equation}
   t_n = t - 2 n T - \sum_{m=0}^{n-1} \delta T_{\mathrm{r.t.}}(t - 2mT) ,
\end{equation}
where $T \equiv L/c$. The corresponding phase can then be written as 
\begin{equation}\label{phiN}
   \phi_n(t) \equiv \omega t_n = 
      \omega t - 2 n k L - \sum_{m=0}^{n-1} \psi(t - 2mT) ,
\end{equation}
where $\psi$ is the phase induced by the gravitational wave in
one round trip (\ref{psiDef}). Substituting (\ref{phiN}) into
(\ref{superPos}), we obtain
\begin{equation}\label{Ecav1}
   {\mathcal{E}}(t) = \sum_{n=1}^{\infty} A_n \rme^{\rmi \omega t} 
     \exp \left[ - 2 \rmi n kL - 
     \rmi \sum_{m=0}^{n-1} \psi(t - 2mT) \right] .
\end{equation}
This formula describes the superposition of an infinite series of
electromagnetic waves in a Fabry-Perot resonator which is placed in
the field of a weak gravitational wave. The phases $\psi$ are
determined by the propagation times of the associated photon round
trips (\ref{psiDef}).

\subsection{Amplification of phase}

We can now identify the amplitude and the phase of the internal
field from (\ref{Ecav1}). First, we notice that $A_n$ contains $(r r')^n$
which can be combined with $\exp(-2 \rmi n kL)$ to yield $\rho^n$, where 
$\rho = r r' \exp(-2 \rmi kL)$. Second, we notice that $\psi$ 
is very small (on the order of $h$) and therefore 
we can approximate: $\rme^{-\rmi \psi}
\approx 1 -\rmi \psi $. With these changes, (\ref{Ecav1}) becomes
\begin{equation}\label{Ecav2}
   {\mathcal{E}}(t) = - \frac{\tau}{r} A_0 
      \rme^{\rmi \omega t} \left( 
      \sum_{n=1}^{\infty} \rho^n - \rmi \sum_{n=1}^{\infty} 
      \sum_{m=0}^{n-1} \gamma_{nm} \right) ,
\end{equation}
where we introduced the temporary notation:  
$\gamma_{nm} = \rho^n \psi(t - 2mT)$. 
The first sum in (\ref{Ecav2}) is a geometric progression which
amounts to $\rho/(1-\rho)$. The second (double) sum in (\ref{Ecav2}) can
be simplified by resummation:
\begin{equation}
   \sum_{n=1}^{\infty} \sum_{m=0}^{n-1} \gamma_{nm} =
   \sum_{m=0}^{\infty} \sum_{n=m+1}^{\infty} \gamma_{nm} = 
   \frac{\rho}{1 - \rho} \sum_{m=0}^{\infty} \rho^{m} \psi(t - 2mT) .
\end{equation}
With these changes, (\ref{Ecav2}) becomes
\begin{equation}
   {\mathcal{E}}(t) = - \frac{\tau}{r} A_0 
   \rme^{\rmi \omega t} \frac{\rho}{1 - \rho} 
   \left[ 1 - \rmi \sum_{m=0}^{\infty} \rho^{m} \psi(t - 2mT) \right] .
\end{equation}
To first order in $h$, the expression in the square brackets can be
approximated by a phase factor $\rme^{-\rmi \Psi(t)}$. Therefore,
\begin{equation}
   {\mathcal{E}}(t) = A \, \rme^{\rmi \omega t - \rmi \Psi(t)} ,
\end{equation}
where $A$ is the unperturbed amplitude of the field and $\Psi(t)$ is
the phase shift due to the gravitational wave. In the explicit form, 
\begin{equation}\label{Aref}
   A = - \frac{\tau \rho}{r(1 - \rho)} A_0 
\end{equation}
and 
\begin{equation}\label{PsiSum}
   \Psi(t) = \sum_{m=0}^{\infty} \rho^{m} \psi(t - 2mT) .
\end{equation}
The Laplace-domain version of (\ref{PsiSum}) can be written as
\begin{equation}\label{PsiCofS}
   \tilde{\Psi}(s) = g_0 C(s) \tilde{\psi}(s) ,
   \qquad
   C(s) = \frac{1 - \rho}{1 - \rho \, \rme^{-2 s T}} ,
\end{equation}
where $g_0 = (1 - \rho)^{-1}$ is the amplification factor and $C(s)$ 
is the transfer function of a Fabry-Perot resonator normalized to $1$ 
at $s=0$. In a perfectly tuned Fabry-Perot resonator, the length $L$ 
is equal to an integer number of laser half-wavelengths and therefore 
$\rme^{-2\rmi kL}=1$ and $\rho = r r'$. We have thus derived equations 
(26) and (27) in \cite{Rakhmanov:2008}. Note that the 4-km LIGO 
interferometers have gain $g_0 \approx 70$ which amounts to a 70-fold 
increase in the sensitivity of these detectors from the Fabry-Perot 
effect.

\section{Response of detectors to gravitational waves}

We conclude with a brief derivation of the detector response functions 
for two widely known configurations: a simple Michelson interferometer 
and a Michelson interferometer with Fabry-Perot arm cavities. For a 
schematic diagram of these detectors and definitions of the field 
amplitudes see figure 2 in \cite{Rakhmanov:2008}.

In a simple Michelson configuration, the signal is produced by 
interfering two beams which return to the beam splitter after 
completing a round trip in the interferometer arms. Denote the phases 
induced by the gravitational wave in each arm by $\psi(t,{\hat{a}})$ 
and $\psi(t,{\hat{b}})$, where $\hat{a}$ and $\hat{b}$ are the unit 
vectors along the interferometer arms. Once the unit vectors are 
defined, the phases can be found from (\ref{psiDef}) and 
(\ref{TrtViaC}). If the beam splitter is set to the dark fringe, 
the signal is proportional to the difference in the phases of 
the two beams. With appropriate normalization 
\cite{Rakhmanov:2008}, it is given by 
\begin{equation}
   V(t) = \frac{1}{2 \omega T} \left[ 
          \psi(t, {\hat{a}}) - \psi(t, {\hat{b}}) \right].
\end{equation}
Taking the Laplace transformation and substituting for the phases from 
(\ref{psiDef}) and (\ref{TrtDs}), we obtain
\begin{equation}
   \tilde{V}(s) = G_{+}(s) \tilde{h}_{+}(s) + 
   G_{\times}(s) \tilde{h}_{\times}(s) ,
\end{equation}
where $G_{i}(s)$ are the transfer functions for the two polarizations 
of the gravitational wave:
\begin{eqnarray}
   G_{+}(s) & = & \frac{1}{2} (a_x^2 - a_y^2) D(s, {\hat{a}}) - 
                  \frac{1}{2} (b_x^2 - b_y^2) D(s, {\hat{b}}) , 
                  \label{eqForGp}  \\
   G_{\times}(s) & = & a_x a_y D(s, {\hat{a}}) - 
                      b_x b_y D(s, {\hat{b}}) .
                      \label{eqForGc}
\end{eqnarray}
Equations (\ref{eqForGp}) and (\ref{eqForGc}) are equivalent to 
equation (19) in \cite{Rakhmanov:2008}.

Similarly, in a Michelson configuration with Fabry-Perot resonators
in the interferometer arms, the signal is given by 
\begin{equation}
   V(t) = \frac{1}{2 \omega T} \left[ 
          \Psi(t, {\hat{a}}) - \Psi(t, {\hat{b}}) \right] ,
\end{equation}
where the phases $\Psi(t,{\hat{a}})$ and $\Psi(t,{\hat{b}})$
correspond to the internal fields in each resonator arm. They can be 
found from (\ref{PsiSum}). In the Laplace domain, the signal can be
written as
\begin{equation}
   \tilde{V}(s) = g_0 \left[ H_{+}(s) \tilde{h}_{+}(s) + 
   H_{\times}(s) \tilde{h}_{\times}(s) \right] .
\end{equation}
The corresponding transfer functions are 
\begin{equation}
   H_{i}(s) = C(s) \, G_{i}(s) ,
\end{equation}
where  $C(s)$ is given by (\ref{PsiCofS}) and $G_{i}(s)$ are given by 
(\ref{eqForGp}) and (\ref{eqForGc}). We have thus derived equations 
(30) and (31) in \cite{Rakhmanov:2008}. (For normalization purpose, 
the factor $g_0$ was ommitted in \cite{Rakhmanov:2008}.)

\section{Conclusion}

We have seen that the response of laser interferometers to 
gravitational waves derived within the framework of general 
relativity agrees with that of the traditional semi-rigorous approach. 
The present analysis provides a rigorous derivation of the key 
formulae in \cite{Rakhmanov:2008} which constitute an accurate model 
for the detector response valid at all frequencies. Such an accurate 
model is required for analysis of the detector's sensitivity at high 
frequencies where the conventional long-wavelength approximation 
breaks down. This includes recent studies of the possibility of 
searches for gravitational waves at the free spectral range (37.5~kHz) 
of the 4-km LIGO interferometers \cite{Sigg:2003, Hunter:2005}, where 
the sensitivity peaks in a narrow ($\sim 200$-Hz) band. The increased 
sensitivity of the detectors at this frequency led to the development 
of new data-analysis efforts targeting searches for high-frequency 
burst \cite{Parker:2007} and stochastic \cite{Giampanis:2008} 
gravitational waves.

\ack

The author acknowledges the support from the Center for Gravitational 
Wave Physics at the Pennsylvania State University and the University 
of Texas at Brownsville. This work was supported by the US National 
Science Foundation under grant PHY 0734800. This paper has been 
assigned LIGO Document Number P070060 and was circulated for the 
internal review by the LIGO Scientific Collaboration on Oct. 23, 2008.

\section*{References}

\providecommand{\newblock}{}

\end{document}